# Anticipation of summer monsoon rainfall over India by Artificial Neural Network with Conjugate Gradient Descent Learning

Surajit Chattopadhyay

Department of Information Technology

Pailan College of Management and Technology

Affiliated to West Bengal University of Technology

Bengal Pailan Park

Kolkata 700 104

West Bengal

E-mail:surajit_2008@yahoo.co.in

## Abstract

Present study aims to develop a predictive model for the average summer monsoon rainfall amount over India. The dataset made available by the Indian Institute of Tropical Meteorology, Pune, has been explored. To develop the predictive model, Backpropagation method with Conjugate Gradient Descent algorithm has been implemented. The Neural Net model with the said algorithm has been learned thrice to reach a good result. After three runs of the model, it is found that a high prediction yield is available. Ultimately, Artificial Neural Network with Conjugate Gradient Descent based Backpropagation algorithm is found to be skillful in predicting average summer monsoon rainfall amount over India.

## Introduction

Artificial neural networks (ANN) are parallel computational models; comprising closely interconnected adaptive processing units. The important characteristic of neural networks is their adaptive nature, where 'learning by example replaces programming'. This feature makes the ANN techniques very appealing in application domains for solving highly non-linear phenomena. During last four decades, various complex problems like weather prediction, stock market prediction etc has been proved to be areas with ample scope of application of this sophisticated mathematical tool. A multilayer neural network can approximate any smooth, measurable function between input and output vectors by selecting a suitable set of connecting weights and transfer functions.

Weather forecasting is one of the most urgent and challenging operational responsibilities carried out by meteorological services all over the world. It is a complicated procedure that includes numerous specialized fields of know-how [4]. Several authors (i.e. [1], [3], [26], [27] and many others) have discussed the uncertainty associated with the weather systems. Chaotic features associated with the atmospheric phenomena also have attracted the attention of the modern scientists (Refs [17], [22], [23], [24]). Different scientists over the globe have developed stochastic weather models which are basically statistical models that can be used as random number generators whose output resembles the weather data to which they have been fit [27].

Amongst all weather happenings, rainfall plays the most crucial role in human life. Human civilization to a great degree depends upon its frequency and amount to various scales ([4], [6]). Several stochastic models have been attempted to forecast the occurrence of rainfall, to investigate its seasonal variability, to forecast monthly/ yearly rainfall over some given geographical area. Daily precipitation occurrence has been viewed through Markov chain by Chin, 1977 [2]. Gregory et al (1993) [3] applied a chain-dependent stochastic model, named as Markov chain model to investigate inter annual variability of area average total precipitation. Wilks (1998) [27] applied mixed exponential distribution to simulate precipitation amount at multiple sites exhibiting realistic spatial correlation. Hu (1964) [10] initiated the implementation of ANN in weather forecasting. Since the last few decades, voluminous development in the application field of ANN has opened up new avenues to the forecasting task involving atmosphere related phenomena (Refs [7], [9]). Michaelides et al (1995) [16] compared the performance of ANN with multiple linear regressions in estimating missing rainfall data over Cyprus. Kalogirou et al (1997) [12] implemented ANN to reconstruct the rainfall time series over Cyprus. Lee et al (1998) [15] applied Artificial Neural Network in rainfall prediction by splitting the available data into homogeneous subpopulations. Wong et al (1999) [27] constructed fuzzy rule bases with the aid of SOM and Backpropagation neural networks and then with the help of the rule base developed predictive model for rainfall over Switzerland using spatial interpolation.

Despite so much of emphasis given to the application of ANN in prediction of different weather events all over the globe, Indian meteorological forecasters did not put much precedence on the application this potent mathematical tool in atmospheric prediction. Author of the present papers thinks that summer monsoon rainfall, a highly influential weather phenomenon in Indian agro economy, may be an area that can be immensely developed with application of ANN. After a thorough literature survey, this author found only one application by Guhathakurta (2006) [4] in predicting Indian summer monsoon rainfall. But, Guhathakurta (2006) [4] confined his study within a state of India. Present contribution deviates from the study of Guhathakurta (2006) [4] in the sense that instead of choosing a particular state; the author implements ANN to forecast the average summer-monsoon rainfall over the whole country.

## Data analysis and problem description

This paper develops ANN model step-by-step to predict the average rainfall over India during summer- monsoon by exploring the data available at the website http://www.tropmet.res.in published by Indian Institute of Tropical Meteorology. The problem discussed in this paper is basically a predictive problem. In this paper, four predictors have been used with one predictand. The four predictors are

- Homogenized Indian rainfall in June
- Homogenized Indian rainfall in July
- Homogenized Indian rainfall in August
- Homogenized average summer-monsoon rainfall in India

All these four predictors are considered for the year $Y$ to predict the average summer-monsoon rainfall in India in the year ($Y$+1). The whole data set comprises the rainfall data of the years 1871-1999. The detailed statistics of the data are given in Table 01. This table shows high values of variance showing chaotic nature dataset. Thus, ANN is found to be a suitable method for handling this dataset.

First step towards development of an ANN model is to divide the whole dataset into training and test sets. The training data are taken from 1871 to 1971. The test data are taken from 1972 to 1999.

## Methodology

In the present paper, an ANN based predictive model is developed for the homogenized average monsoon rainfall using the Backpropagation learning through the method of conjugate gradient descent. The Backpropagation learning is based on the gradient descent along the error surface ([5], [11], [14]). That is, the weight adjustment is proportional to the negative gradient of the error with respect to the weight. In mathematical words [14]

$$w_{k+1} = w_k + \eta d_k \qquad \ldots \qquad \ldots \qquad \ldots \qquad (1)$$

Where, $w_l$ denotes the weight matrix at epoch *l*. the positive constant $\eta$, which is selected by the user, is called the learning rate.

The direction vector $d_k$ is negative of the gradient of the output error function *E*

$$d_k = -\nabla E(w_k) \qquad \ldots \qquad \ldots \qquad \ldots \qquad (2)$$

There are two standard learning schemes for the BP algorithm: on-line learning and batch learning. In on-line learning, the weights of the network are updated immediately after the presentation of each pair of input and target patterns. In batch learning all the pairs of patterns in the training sets are treated as a batch, and the network is updated after processing of all training patterns in the batch. In either case the vector $w_k$ contains the weights computed during *k*th iteration, and the output error function *E* is a multivariate function of the weights in the network [14]:

$$E(w_k) = \begin{cases} E_p(w_k)[on-line] \\ \sum_p E_p(w_k)[Batch] \end{cases} \qquad \ldots \qquad \ldots \qquad \ldots \qquad (3)$$

Where, $E_p(w_k)$ denotes the half - sum – of – squares error functions of the network output for a certain input pattern *p*. The purpose of the supervised learning (or training) is to find out a set of weight that can minimize the error *E* over the complete set of training pair. Every cycle in which each one of the training patterns is presented once to the neural network is called an epoch (for details see [13], [18], [19], [20], [29]).

The direction vector $d_k$, expressed in terms of error gradient depends upon the choice of activation function. When the sigmoid function i.e. $f(x) = (1+\exp(-x))^{-1}$ is adopted, the BP algorithm becomes 'Back propagation for the Sigmoid Adaline' [28]

Normally, the Backpropagation learning uses the weight change proportional to the negative gradient of the instantaneous error. Thus it uses the only the first derivative of the instantaneous error with respect to the weight. A more effective method [29] can be derived by starting with the following Taylor series expansion of the error as a function of the weight vector

$$E(w+\Delta w) = E(w) + g^T \Delta w + \frac{1}{2}\Delta w^T H \Delta w + ................\ldots \qquad \ldots \qquad (4)$$

Where, $g = \frac{\Delta E}{\Delta w}$ is the gradient vector, and $H = \frac{\partial^2 w}{\partial w^2}$ the Hessian matrix. Newton's method of Backpropagation learning uses this Hessian matrix as a component of weight update.

Conjugate Gradient Descent method, adopted in the present paper, provides a better learning algorithm for Backpropagation method. In this method, the increment of the weight at mth step is given by [29]:

$$\Delta w = w(m+1) - w(m) = \eta(m) d(m) \ldots \qquad \ldots \qquad \ldots \qquad (5)$$

Where the direction of the increment $d(m)$ in the weight is a linear combination of the current gradient vector and the previous direction of the increment in the weight. That is

$$d(m) = -g(m) + \alpha(m-1) d(m-1) \qquad \ldots \qquad \ldots \qquad \ldots \qquad (6)$$

Where, $\alpha(m) = \frac{g^T(m+1)[g(m+1) - g(m)]}{g^T(m) g(m)} \qquad \ldots \qquad \ldots \qquad (7)$

## Implementation and results

The available data set are first scaled according to

$$z_i = 0.1 + 0.8 \times \left( \frac{x_i - x_{\min}}{x_{\max} - x_{\min}} \right) \qquad \ldots \qquad \ldots \qquad (8)$$

Where, $z_i$ denotes the transformed appearance of the raw data $x_i$.

After the modeling is completed, the scaled data are reverse scaled according to

$$P_i = x_{\min} + \left( \frac{1}{0.8} \right) \times \left[ (y_i - 0.2) \times (x_{\max} - x_{\min}) \right] \qquad \ldots \qquad \ldots \qquad (9)$$

Where, $P_i$ denotes the prediction in original scale, and the corresponding scaled prediction is $y_i$.

The whole dataset is then divided into training and test sets. The first 75% of the whole dataset is taken as the training set and the last 25% of the data are taken as the test set. Using Conjugate Gradient Descent algorithm, the data are trained three times up to 1000 epochs. After training the ANN is tested over the test set. Evolution of mean squared error during the training process (three runs) is presented in Figure 01. Table 02 depicts the test year, the predictor values and the predicted values of average summer monsoon rainfall over India. The actual and predicted summer monsoons in the test years are presented in Figure 02. The figure shows a close association between actual rainfall amounts and those predicted by ANN with Conjugate Gradient Descent Backpropagation. In Figure 03, the error of prediction (%) in each test case is presented. It is observed that in 78% of the test cases, the prediction error lies below 20% and in 61% cases the prediction error lies below 10%. This proves a high prediction yield by the Backpropagation ANN with Conjugate Gradient descent learning.

## Conclusion

The study proves the nimbleness of ANN as a predictive tool for average summer monsoon rainfall in India. Furthermore, Conjugate Gradient Descent is proved to be an efficient Backpropagation algorithm that can be adopted to predict the average summer monsoon rainfall time series over India.

Table 01-Detailed statistical properties of the time series pertaining to average rainfall over India in the summer monsoon months

| | **June** | **July** | **August** |
|---|---|---|---|
| Mean | 127.5705426 | 259.4666667 | 223.4472868 |
| Median | 128.5 | 256.1 | 221 |
| Mode | 177.1 | 251.6 | 168.6 |
| Standard Deviation | 43.73563586 | 56.81155711 | 59.52120956 |
| Sample Variance | 1912.805844 | 3227.553021 | 3542.774387 |
| Kurtosis | -0.582420137 | 0.3709231 | -0.762070931 |
| Skewness | 0.036895646 | -0.316501162 | 0.159888472 |
| Range | 196.8 | 294.2 | 256.7 |
| Minimum | 30 | 94.5 | 88.8 |
| Maximum | 226.8 | 388.7 | 345.5 |
| Sum | 16456.6 | 33471.2 | 28824.7 |
| Count | 129 | 129 | 129 |
| Largest(1) | 226.8 | 388.7 | 345.5 |
| Smallest(1) | 30 | 94.5 | 88.8 |

Table 02- Tabular presentation of the predictors and the predictands in the test years

| Test Year (Y+1) | June rain year Y (mm) | July rain year Y (mm) | August rain year Y (mm) | Average rain year Y (mm) | Average rain year (Y+1) (mm) | ANN Predicted rain year (Y+1) (mm) |
|---|---|---|---|---|---|---|
| **1972** | 196.1 | 225.5 | 210.6 | 210.7333333 | 154.2333333 | 200.9549209 |
| **1973** | 96.8 | 150.1 | 215.8 | 154.2333333 | 237.9666667 | 234.3711035 |
| **1974** | 85.3 | 294.8 | 333.8 | 237.9666667 | 160.4 | 214.2280936 |
| **1975** | 72.3 | 205.3 | 203.6 | 160.4 | 227.5333333 | 224.769791 |
| **1976** | 149.5 | 260 | 273.1 | 227.5333333 | 228.5666667 | 198.2542725 |
| **1977** | 125.3 | 287.2 | 273.2 | 228.5666667 | 232.0333333 | 201.9250574 |
| **1978** | 178.6 | 291.7 | 225.8 | 232.0333333 | 242.8333333 | 193.6718046 |
| **1979** | 174.3 | 276.7 | 277.5 | 242.8333333 | 177.9666667 | 194.5932308 |
| **1980** | 123.5 | 179.6 | 230.8 | 177.9666667 | 230.3 | 217.9035457 |
| **1981** | 202.1 | 235.8 | 253 | 230.3 | 192.2333333 | 193.933663 |
| **1982** | 111.1 | 253.8 | 211.8 | 192.2333333 | 182.5666667 | 202.1781142 |
| **1983** | 79.5 | 219.1 | 249.1 | 182.5666667 | 229.1 | 213.0060783 |
| **1984** | 118.4 | 266.7 | 302.2 | 229.1 | 192.8333333 | 207.761418 |
| **1985** | 95.4 | 217.5 | 265.6 | 192.8333333 | 176.6 | 209.6381144 |
| **1986** | 97.9 | 230.8 | 201.1 | 176.6 | 193.8333333 | 205.5389094 |
| **1987** | 163.3 | 221.1 | 197.1 | 193.8333333 | 150.6 | 211.4320826 |
| **1988** | 83.4 | 157.8 | 210.6 | 150.6 | 221.2333333 | 236.7569004 |
| **1989** | 115.9 | 321 | 226.8 | 221.2333333 | 205.1666667 | 198.4402201 |
| **1990** | 150.1 | 232.6 | 232.8 | 205.1666667 | 235.4666667 | 200.7099469 |
| **1991** | 155.5 | 227.6 | 323.3 | 235.4666667 | 201.8666667 | 203.9973342 |
| **1992** | 138.8 | 260.8 | 206 | 201.8666667 | 196.6666667 | 201.9776989 |
| **1993** | 94.8 | 187.5 | 307.7 | 196.6666667 | 204.0666667 | 220.7533005 |
| **1994** | 135.1 | 287 | 190.1 | 204.0666667 | 261.7666667 | 202.0665298 |
| **1995** | 206.6 | 322.1 | 256.6 | 261.7666667 | 175.9333333 | 192.1425691 |
| **1996** | 73 | 285.7 | 169.1 | 175.9333333 | 205.0333333 | 191.1645972 |
| **1997** | 142.3 | 243.1 | 229.7 | 205.0333333 | 204.2333333 | 200.3300993 |
| **1998** | 150.3 | 219.2 | 243.2 | 204.2333333 | 183.6 | 201.7475505 |
| **1999** | 131.6 | 219.2 | 200 | 183.6 | 169.6333333 | 209.2529582 |

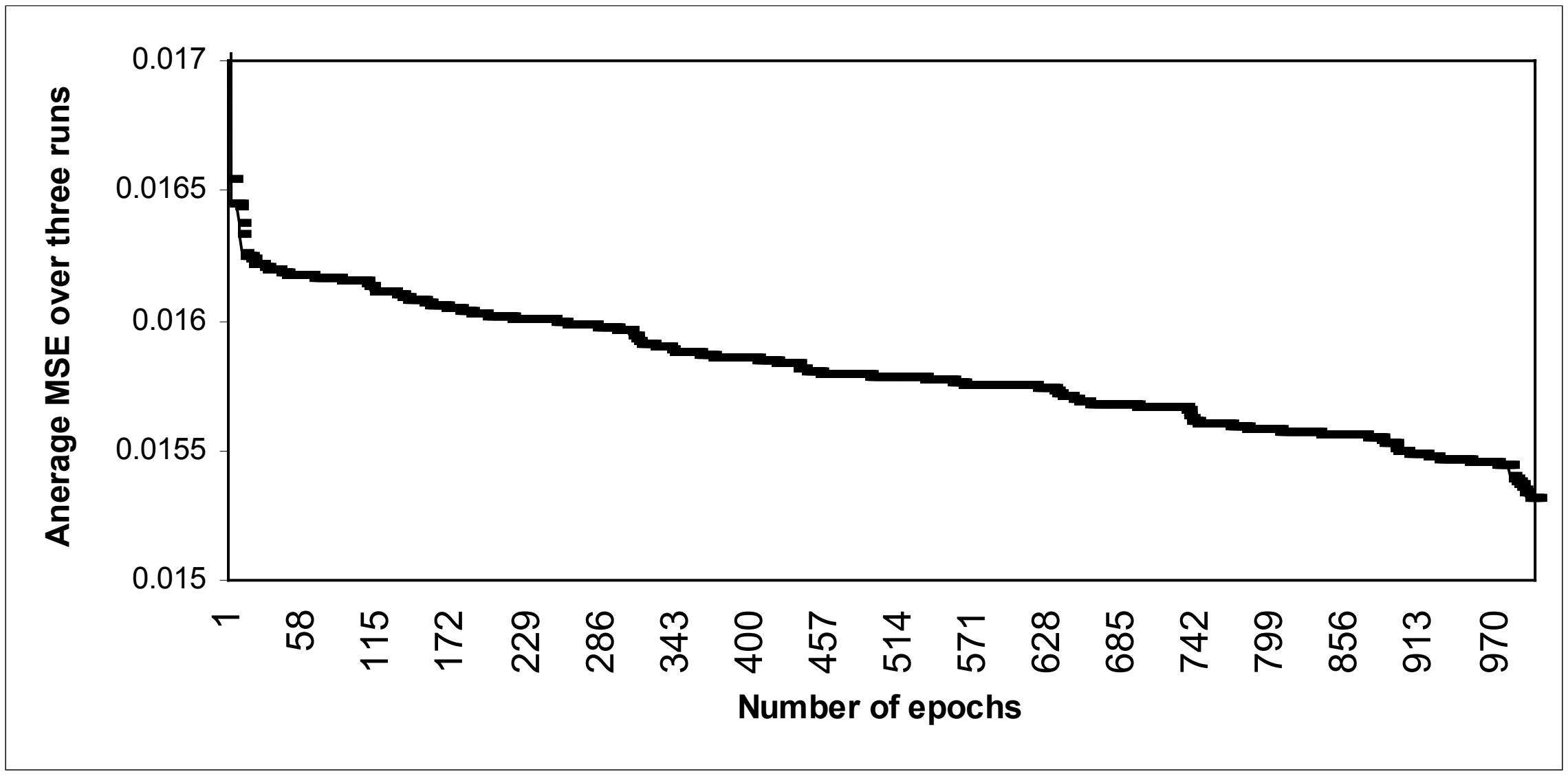

Fig.01- Schematic of the evolution of the mean squared errors (averaged over three runs) with increase in the number of epochs.

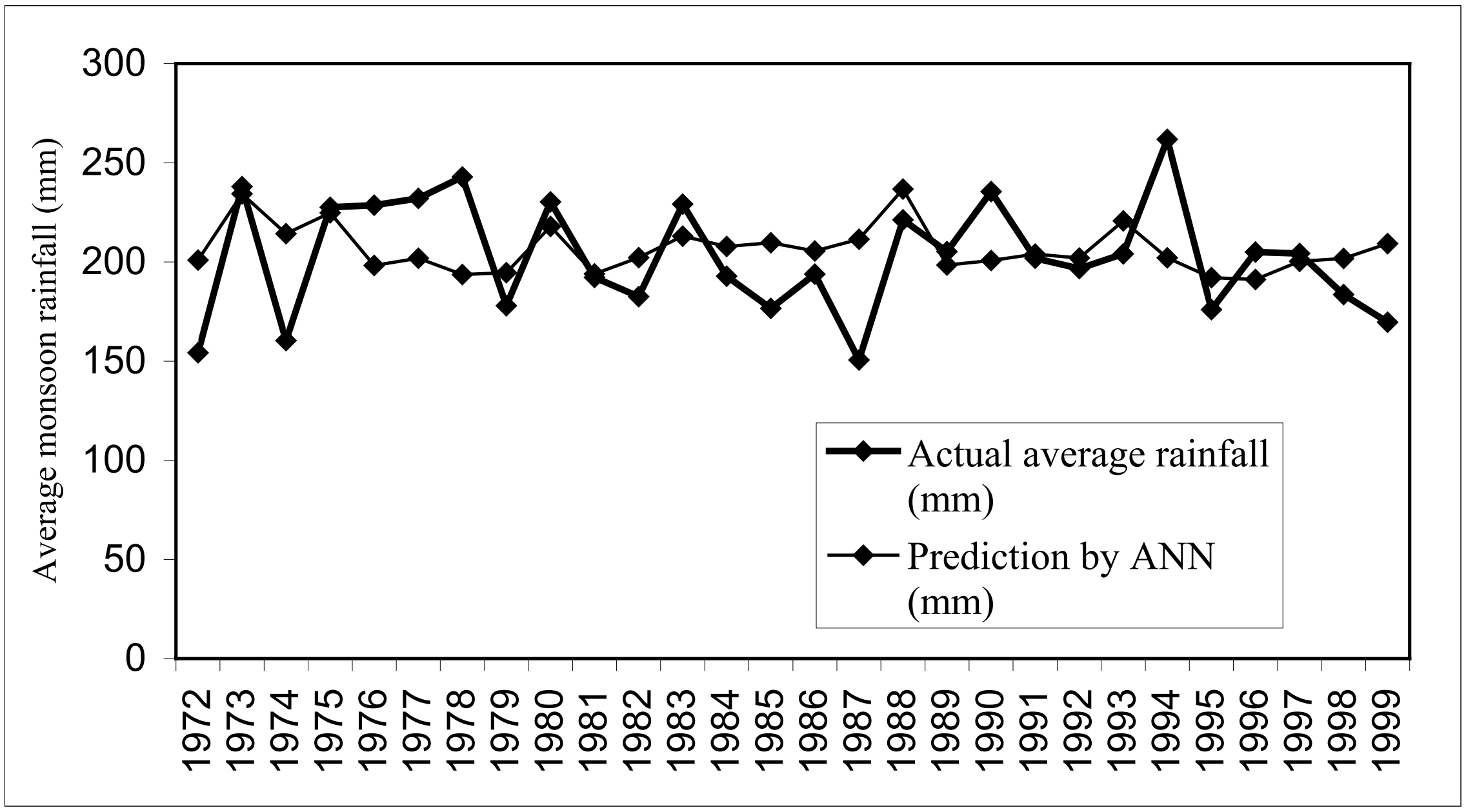

Fig.02- Schematic showing the actual and predicted average monsoon rainfall amounts (mm) in the test years,

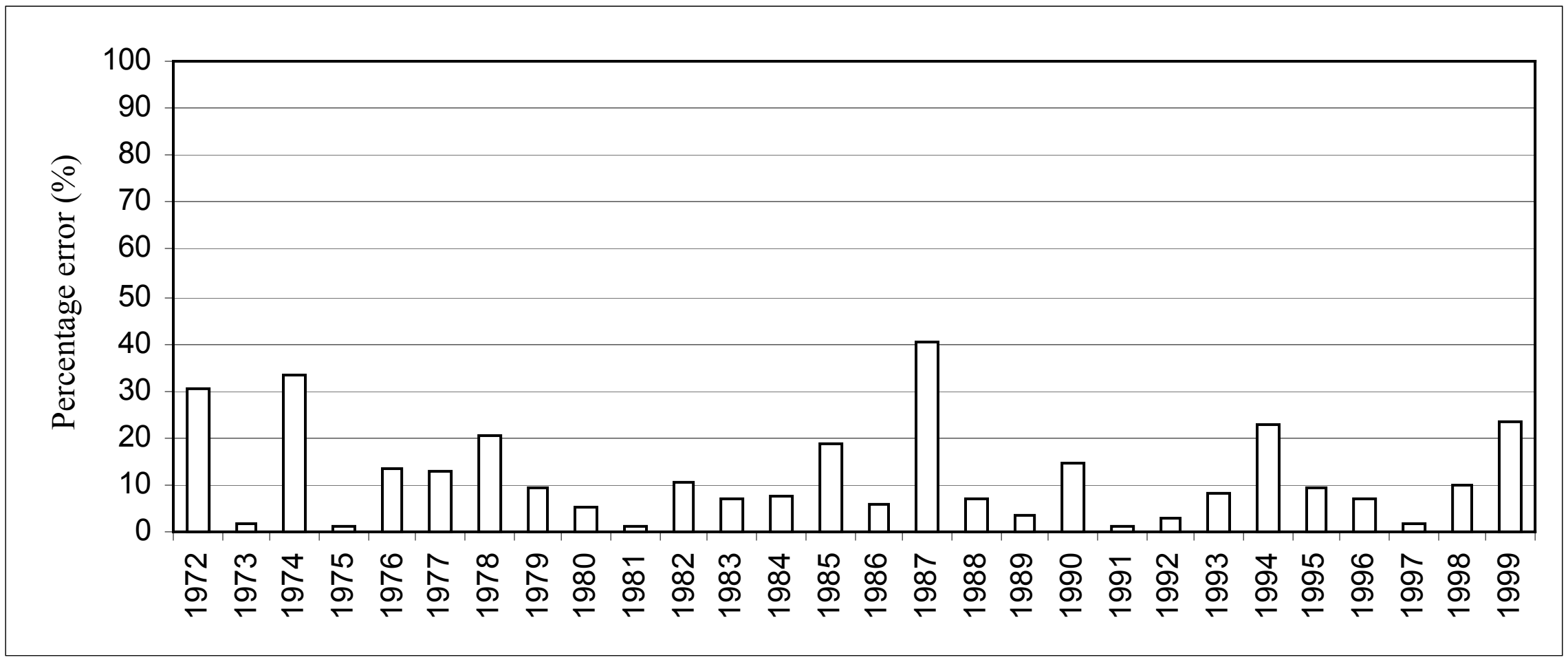

Fig.03- Percentage errors of prediction in the test years.